\documentclass[preprint]{elsarticle}
\usepackage{fullpage}
\usepackage{algorithmic}
\usepackage{mathtools}
\usepackage{hyperref}

\DeclarePairedDelimiter\ceil{\lceil}{\rceil}
\title{Planar lower envelope of monotone polygonal chains}
\journal{Information Processing Letters}
\begin{document}

\begin{frontmatter}
\author{Daniel L. Lu}
\ead{dllu@cmu.edu}
\address{Carnegie Mellon University,
Robotics Institute,
5000 Forbes Ave,
Pittsburgh, PA 15213}
\begin{abstract}
A simple linear search algorithm running in $O(n+mk)$ time is proposed for constructing the lower envelope of $k$ vertices from $m$ monotone polygonal chains in 2D with $n$ vertices in total. This can be applied to output-sensitive construction of lower envelopes for $n$ arbitrary line segments in optimal $O(n\log k)$ time, where $k$ is the output size. Compared to existing output-sensitive algorithms for lower envelopes, this is simpler to implement, does not require complex data structures, and is a constant factor faster.
\end{abstract}
\end{frontmatter}

\section{Introduction}
Finding the lower envelope of line segments in 2D is useful in visibility problems in robotics, facility location, architecture, video games, and computer graphics. The lower envelope of $n$ segments has a complexity bounded by the third order Davenport--Schinzel sequence, which is in $O(n\alpha(n))$ where $\alpha(n)$ is the slow-growing inverse Ackermann function. A worst-case optimal divide-and-conquer algorithm running in $O(n\log n)$ time was proposed by Hershberger \cite{hersh} as a refinement of earlier $O(n\alpha(n)\log n)$ time methods. 

Since a visibility polygon can be considered a lower envelope in polar coordinates centered on the query point, a lower envelope algorithm can be applied to finding visibility polygons. Some visibility polygon problems are posed with assumptions which allow faster or simpler algorithms to be used. 
For a visibility polygon of $n$ segments allowed to intersect at only their endpoints, $O(n\log n)$ time algorithms simpler than Hershberger's algorithm are known, based on either divide-and-conquer or angular plane sweep \cite{asano, suri}. For point visibility in a simple polygon, a linear time algorithm suffices \cite{ea, lee}. For point visibility in a complex polygon with $h$ holes and $n$ vertices, a worst-case optimal $O(n+h\log h)$ time algorithm using Chazelle's linear time triangulation is known \cite{hm}. Some algorithms using preprocessing have been proposed, involving various tradeoffs between preprocessing time, query time, and storage space \cite{bose, aronov, zarei}.

Optimal output-sensitive $O(n\log k)$ time algorithms for lower envelopes, where $k$ is the output size, were proposed by Chan \cite{chan} and Nielsen and Yvinec \cite{nielsen}. Both use Hershberger's algorithm as a subroutine. Parallelized versions of Nielsen and Yvinec's method have been proposed, including a deterministic $O(\log n (\log k + \log \log n))$ time algorithm and faster randomized variants, each doing a total of $O(n\log k)$ work \cite{gupta}.
However, these all require complex data structures such as lazy interval trees. This paper presents a simple extension of the Jarvis march, similar to Welzl's observation in Idea 5 of Chan's paper \cite{chan}, to monotone polygonal chains. Combined with any $O(n\log n)$ time algorithm, this allows a simpler implementation of Chan's algorithm running in deterministic $O(n\log k)$ time without preprocessing or parallelization.

Although output-sensitive algorithms for computing lower envelopes have been proposed, no known implementation exists; however, efficient implementations of $O(n\log n)$ time visibility polygon algorithms are available in the Computational Geometry Algorithms Library (CGAL) \cite{bungiu} as well as elsewhere \cite{byron}. These can be easily adapted using my algorithm to obtain an optimal $O(n\log k)$ running time. 
An implementation of a divide-and-conquer lower envelope algorithm running in $O(n\alpha(n)\log n)$ time is also available in the CGAL \cite{ronwein}. Using this instead of an $O(n\log n)$ time subroutine yields an overall $O(n\alpha(k) \log k)$ time complexity.

\section{Lower envelope of monotone polygonal chains}
Suppose there are $m$ polygonal chains $V_0,\ldots, V_{m-1}$ with $n$ vertices in total, where the $j$th vertex of the $i$th chain is denoted $V_i(j) = (V_i(j)_x, V_i(j)_y)$ such that $V_i(j)_x \leq V_i(j+1)_x$ and, without loss of generality, $V_i(0)_x=0$ and $V_i(|V_i|-1)_x=1$. For simplicity, we assume no vertex lies on a segment except at its endpoints, and no three segments intersect at a single point, otherwise slopes would have to be compared in addition to positions. A lower envelope is found by following the \textit{active} chain, denoted $V_a$, until another chain intersects it. To efficiently find such intersections, we maintain a pointer $p_i$ on the chain $V_i$, which only advances but never retreats.

On each iteration, a vertex $v^*$ is added to the lower envelope being constructed. Consider the \textit{active segment} $S^* = (v^*, V_a(p_a))$, where $v^*$ lies on the segment $(V_a(p_a-1), V_a(p_a))$. On the first iteration, $S^*$ is initialized to the first segment of $V_a$.
An invariant is that $v^*$ is not occluded. During an iteration, for each $i=0,\ldots,m-1,~i\neq a$, the pointer $p_i$ is incremented until at least one of the following three conditions is met: $V_i(p_i)_x > V_a(p_a)_x$; $S^*$ intersects the segment $S_i = (V_i(p_i-1), V_i(p_i))$ at a point $I \neq v^*$; or $S^*$ intersects the segment $(V_i(p_i-1), (V_i(p_i-1)_x, +\infty))$.
That is, $p_i$ is incremented until $S_i$ is no longer occluded by a combination of $S^*$ and all previously active segments. While each $p_i$ is being incremented, the intersection of $S_i$ and $S^*$ at a point $I\neq v^*$ with the smallest $x$-coordinate is found (if $S^*$ is vertical, the intersection with the smallest $y$-coordinate is found instead). At the end of the iteration, if such an intersection exists, the next $S^*$ is set to the segment from the intersection point to the end of the intersecting segment; otherwise $p_a$ is incremented, and the next $S^*$ is set to be $(V_a(p_a-1), V_a(p_a))$. This is repeated until the entire envelope has been found, which happens when $v^*$ is the last vertex on a chain. A pseudocode is shown in the Appendix.

The total number of increments of pointers $p_i$ is $O(n)$ since they never retreat; the comparisons on each increment are done in constant time. Each time a new vertex is added to the output, the algorithm loops over all $m$ polygonal chains to check for intersection with the currently examined edge and to test if the associated pointer needs to be incremented. Since the output has $k$ vertices, this takes $O(mk)$ time. The total time complexity is thus $O(n + mk)$. The algorithm adapts easily
to polar coordinates to find the intersection of star-shaped polygons (visibility polygons) or the lower envelope of arbitrary piecewise functions composed of pieces which intersect at most once.

\section{Output-sensitive construction of lower envelopes}
Here, a straightforward application of Chan's algorithm \cite{chan} is described.
\begin{algorithmic}[1]
    \FOR{$t=0,1,2,\ldots$}
        \STATE $\kappa \leftarrow 2^{2^t}$
        \STATE Arbitrarily partition segments into $\ceil*{\frac{n}{\kappa}}$ subsets each of size at most $\kappa$
        \STATE Run any $O(n\log n)$ time algorithm on each group, yielding $\ceil*{\frac{n}{\kappa}}$ monotone polygonal chains
        \STATE Find the lower envelope of these monotone polygonal chains, and abort if the output size exceeds $\kappa$
    \ENDFOR
\end{algorithmic}

The time complexity is analyzed here per iteration:
\begin{itemize}
\item Running a $O(n\log n)$ time algorithm (such as Hershberger's algorithm \cite{hersh}, or, for non-intersecting segments, one of the algorithms implemented in \cite{bungiu}) on each group takes $O(\kappa\log\kappa)$ time each, for a total of $O(n\log\kappa)$. The number of vertices in each of the $\ceil*{\frac{n}{\kappa}}$ chains is in $O(\kappa\alpha(\kappa))$.
\item Finding the lower envelope takes $O(n\alpha(\kappa))$ time since the total number of vertices in the $\ceil*{\frac{n}{\kappa}}$ chains is $O(n\alpha(\kappa))$ and the algorithm immediately aborts when the output exceeds size $\kappa$.
\end{itemize}
In particular, the second term of time complexity $O(n\alpha(\kappa))$ is nearly a log factor improvement upon Chan's ray-shooting method \cite{chan}. Ultimately each iteration runs in $O(n\log\kappa)=O(n2^t)$, dominated by the first term. The total time complexity of the $\ceil*{\log\log k}$ iterations is as desired:
\begin{align}
O\left(\sum_{t=1}^{\ceil{\log\log k}} n2^t\right) = O\left(n2^{\ceil{\log\log k}+1}\right) = O(n\log k).\nonumber
\end{align}
It is easy to see that, if the $O(n\log n)$ time algorithm used in step 3 was replaced with one running in $O(n\alpha(n)\log n)$ time such as \cite{ronwein}, then the overall time complexity increases to $O(n\alpha(k) \log k)$.

\newpage
\section*{Appendix: Pseudocode}
Here is a pseudocode description of the linear search algorithm for finding the lower envelope of monotone polygonal chains $V_0,\ldots, V_{m-1}$. An animated explanation can be found at \\
\texttt{http://www.dllu.net/present\_visibility/envelope.html} with a C++11 implementation at \\
\texttt{http://www.dllu.net/present\_visibility/implementation.cpp}
\begin{algorithmic}[1]
    \STATE $p_0,\ldots, p_{m-1} \leftarrow 1$
    \STATE $a \leftarrow$ the index of the chain with the smallest $y$-coordinate of the first vertex
    \STATE $p_a \leftarrow 1$
    \STATE $v^* \leftarrow V_a(0)$
    \STATE $V \leftarrow \lbrace v^*\rbrace$
    \REPEAT
        \STATE $\operatorname{best} \leftarrow (\infty, \infty)$, $\operatorname{next} \leftarrow -1$
        \FOR{$i=0,\ldots, m-1,~i\neq a$}
        \WHILE{$p_i < |V_i|$ and $V_i(p_i)_x \leq V_a(p_a)_x$ and neither $(V_i(p_i-1), V_i(p_i))$ nor $(V_i(p_i-1), (V_i(p_i-1)_x, +\infty))$ intersects $(v^*, V_a(p_a))$ at a point $I\neq v^*$}
                \STATE $p_i\leftarrow p_i+1$
            \ENDWHILE
            \IF{$(V_i(p_i-1),V_i(p_i))$ intersects $(v^*, V_a(p_a))$ at a point $I$ such that $I_x < \operatorname{best}_x$, or ($I_x = \operatorname{best}_x$ and $I_y < \operatorname{best}_y$)}
                \STATE $\operatorname{best} \leftarrow I$, $\operatorname{next} \leftarrow i$
            \ENDIF
        \ENDFOR
        \IF{$\operatorname{next}=-1$}
            \STATE $v^* \leftarrow V_a(p_a)$
            \STATE $p_a \leftarrow p_a+1$
        \ELSE
            \STATE $v^* \leftarrow \operatorname{best}$
            \STATE $a \leftarrow \operatorname{next}$
        \ENDIF
        \STATE Append $v^*$ to $V$
    \UNTIL{$p_a \geq |V_a|$}
    \STATE \textbf{return} $V$
\end{algorithmic}
\end{document}